\documentclass[aps,pre,twocolumn,showpacs,longbibliography]{revtex4-1} 
\usepackage{amssymb}
\usepackage{amsfonts}
\usepackage{amsmath}
\usepackage{amstext}
\usepackage{braket}
\usepackage{graphicx}
\usepackage{bm}
\usepackage{soul}
\usepackage{xcolor}
\usepackage{marginnote}
\usepackage[normalem]{ulem}
\usepackage{enumitem}
\usepackage{empheq}
\usepackage{bbold}
\usepackage{dsfont}
\providecommand{\U}[1]{\protect\rule{.1in}{.1in}}

\begin{document}
\title{Co-evolution of nodes and links: \\ 
diversity driven coexistence in cyclic
competition of three species}
\author{Kevin E. Bassler}
\email{bassler@uh.edu}
\affiliation{Department of Physics, Department of Physics, University of Houston, Houston,
TX 77204-5005, USA}
\affiliation{Texas Center for Superconductivity, University of Houston, Houston, TX
77204-5002, USA}
\affiliation{Max-Planck-Institut f\"{u}r Physik komplexer Systeme, N\"{o}thnitzer Str. 38,
Dresden D-01187, Germany}
\author{Erwin Frey}
\email{frey@lmu.de}
\affiliation{Arnold Sommerfeld Center for Theoretical Physics and Center for Nanoscience,
Department of Physics, Ludwig-Maximilians-Universit\"{a}t M\"{u}nchen, Theresienstrasse 37, 80333 M\"{u}nchen, Germany}
\author{R. K. P. Zia}
\email{rkpzia@vt.edu}
\affiliation{Max-Planck-Institut f\"{u}r Physik komplexer Systeme, N\"{o}thnitzer Str. 38, Dresden D-01187, Germany}
\affiliation{Center for Soft Matter and Biological Physics, Department of Physics, Virginia Polytechnic Institute and State University, Blacksburg, VA 24061, USA}
\date{\today}

\begin{abstract}
When three species compete cyclically in a well-mixed, stochastic system of $N$ individuals, extinction is known to typically occur at times scaling as the system size $N$. This happens, for example, in rock-paper-scissors games or conserved Lotka-Volterra models in which every pair of individuals can interact on a complete graph. Here we show that if the competing individuals also have a \textquotedblleft social temperament\textquotedblright\ to be either introverted or extroverted, leading them to cut or add links respectively, then long-living state in which all species coexist can occur when both introverts and extroverts are present. These states are non-equilibrium quasi-steady states, maintained by a subtle balance between species competition and network dynamcis. Remarkably, much of the phenomena is embodied in a mean-field description. However, an \textit{intuitive} understanding of why diversity stabilizes the co-evolving node and link dynamics remains an open issue.
\end{abstract}
\maketitle



\section{Introduction}

Evolutionary game theory~\cite{smith,Frey2010} considers populations composed of individuals with different strategies or behavioral programs who compete generation after generation in game situations of the same type. A central question is how evolutionary forces like natural selection and mutation shape the time evolution of a population. In particular, one is interested to learn about mechanisms underlying maintenance of species diversity and extinction of species. Typically one studies fixed environments which may be, for example, well-mixed or spatially extended systems, where the players of the game are located at the nodes of a complete graph or a regular lattice, respectively. On another front, much of network science~\cite{Newman2003,Barabasi2016,Latora2017} has been devoted to the study of the structure or topology of the links in a graph, while the nodes have no degrees of freedom. Though the properties of static networks, such as Erd\H{o}s-R\'{e}nyi random graphs~\cite{ER1959}, have received considerable attention, there has also been interest in networks with links that evolve dynamically, such as in the formation of scale-free networks by preferential attachment~\cite{Price1965,BA1999}. Bringing together these two paradigms, statistical mechanics of node degrees of freedom connected by fixed links and networks with an evolving link topology, is the central motivation of our work. Clearly, in many complex systems in nature, the evolution of both components must be accounted for simultaneously. 

Studies of networks that have `co-evolving' nodes and links, also known as `adaptive networks', began to emerge about two decades ago~\cite{PCB2000,LB2006,Gross2008,Perc2010,Sayama2013}. In attempts to model realistic co-evolving systems, complex mathematical structures and serious challenges are encountered. In this context, we introduce a minimal system of node and link degrees of freedom, co-evolving with stochastic rules. Finding unexpected, novel behavior in extensive numerical simulations, we exploit the simplicity of the model to derive tractable mean-field equations and to obtain some analytic results. This study should be regarded as one of a few first steps towards investigating more complex and realistic co-evolving systems.

Our model combines the node dynamics of the well known rock-paper-scissors
game \cite{Hofbauer1998,Frey2010}, 
also known as cyclic Lotka-Volterra model \cite{Dobrinevski2012,Knebel2013,Knebel2015}, 
and the link dynamics in recent studies of
networks with preferred
degrees~\cite{ZLJS2011,ZLS2012,LSZepl2012,LSZ2014,BLSZpre15,BDZ2015}. In the
former, the three `species' compete cyclically and, if $N$ individuals play
stochastically on a complete network, and the typical extinction time scales 
with the population size $N$~\cite{RMF06}. In the latter system, designed to model the actions
of individuals with different `social temperaments' -- introverts and
extroverts -- are considered. The links fluctuate as a randomly chosen
individual cuts or adds connections. Even in a simple population with extreme
temperaments, surprising behavior emerged \cite{LSZepl2012,BLSZpre15,BDZ2015}.
Here, we consider a system with nodes that can be one of three species and can
have one of two temperaments, with specific rules of evolution for both the
nodes and links. When a link between two nodes is absent, they do not
interact, so that the competition between the three species is both `tempered'
and dynamic. In a following, we report simulation results with $N$ up to
$1000$, as well as a theoretic description, based on mean-field approximation.
Specifically, we discover long-living, \textit{quasi-stationary states} (QSS),
persisting for up to $O\left(  10^{12}\right)  $ node changes. For these QSS to occur, nodes
with both temperaments must be present. Thus, we refer to this kind stability
as `diversity driven coexistence.'

In the next section, we specify the setup of our model and its dynamics. 
Results from simulation studies and analysis of a set of mean-field equations 
are presented in a following section. It is natural that
findings of the QSS states raise more interesting questions for future
research. These issues, as well as an outlook for exploring more realistic
models of co-evolution, are discussed in the final section.

\section{Specifications of the model}

Our model of co-evolution consists of merging two of the simplest statistical
systems, each involving node or link dynamics only. The former is designed for
neutral cyclic competition amongst three species. Known as the cyclic
Lotka-Volterra model, it is often portrayed as the game of rock-paper-scissors. 
Here we refer to it here as the $ABC$ model \cite{RMF06,Knebel2013}. 
The latter consists of a network of links, cut or added
by a collection of `introverts' ($I$) and `extroverts' ($E$), the extreme case
of which is the $XIE$ model~\cite{ZLS2012,LSZepl2012}. Thus, we will refer to
the union as the $ABC$-$XIE$ model.

For the $ABC$ component, let us follow the notation of Refs.~\cite{RMF06,BRSFprl09} and
consider the simplest possible scenario: a system with $N$ individuals, each
being one of three `species' ($A$, $B$, $C$), competing cyclically with unit
rates:
\begin{equation}
A+B\rightarrow2A;~~B+C\rightarrow2B;~~C+A\rightarrow2C \, .
\label{ABC-rules}%
\end{equation}
With no spatial structure, the configuration of a `well-mixed' system is
completely specified by the numbers of each ($N_{A,B,C}$) with the total
number $N=N_{A}+N_{B}+N_{C}$ being a constant of motion. In the large $N$ limit, the
evolution is well approximated by a deterministic set of 'chemical rate' equations:
\begin{equation}
\partial_{t}N_{A}=N_{A}N_{B}-N_{C}N_{A}\text{ \ and cyclic} \, ,
\label{Neqn}%
\end{equation}
where $t$ is appropriately normalized (all rates are taken as unity). 
It is well-known that $R\equiv
N_{A}N_{B}N_{C}$ is a constant of motion of these equations, so that the orbits in
configuration space form closed loops and extinction never occurs \cite{Hofbauer1998}. 
In this
sense, the competition is `neutral'.  
However, if stochastic aspects of the
evolution is included, then extinction of two of the species is inevitable for
finite systems, typically in $O\left(  N\right)  $ steps~\cite{RMF06,Dobrinevski2012}. 
Indeed, highly counter-intuitive behavior is found when the competition rates are
unequal \cite{BRSFprl09}. Similarly, many interesting phenomena emerge when
such systems are placed into some spatial structure~\cite{durrett-1997-185, durrett-1998-53, Czaran2002, Kerr2002, Reichenbach2007, Reichenbach2007b, Reichenbach2008, RulandsZielinskiFrey2013, Weber2014}, in which an individual may interact with, say, only nearest
neighbors in a lattice. 

For the $XIE$ component, the simplest
version~\cite{ZLS2012,LSZepl2012,BLSZpre15} consists of a fixed number of
introverts and extroverts, connected by a network with dynamic links that
change as a result of the action of a randomly chosen individual: An $I$ will
cut one of its existing links while an $E$ will add a link to an individual it
is not already connected to. With random sequential update, this simple model
displays an extreme Thouless effect: an extraordinary transition when the
numbers of the subgroups are equal~\cite{BLSZpre15}.

Our interest here is to study the importance of behavioral diversity
in games of cyclic dominance.
To do so we merge the ABC and XIE models, by endowing the individuals of
the $ABC$ model with a temperament, $\tau \in \{I,E\}$. To avoid confusion, we will
denote `species' by $\alpha \in\{A,B,C\}$, so that an individual's state is
given by $\left(  \alpha,\tau\right)  $. In this sense, each node can be one
of six `types': $A_{I,E}$, $B_{I,E}$, and $C_{I,E}$. A system configuration is
specified by $\mathcal{C}$: $\left\{ \left(  \alpha,\tau\right)_{i}%
;a_{ij}\right\}  $, where $i,j \in\{1,...,N\}$ label an individual and $a_{ij}$ are
elements of $\mathbb{A}$, the adjacency matrix describing the presence or
absence of links.\ Thus, $a_{ij}=1/0$ if the link between $i$ and $j$ is
present/absent. There are no self-loops: $a_{ii}\equiv0$.

A related model has studied cyclic dominance on an adaptive network in a model 
of opinion spreading~\cite{Demirel2014}.
However, the link dynamics were very different in that model.
All nodes behaved identically.
Link updates were coupled to the outcome of node interactions
through an ABC competition.
A losing node either chose to adopt the ``opinion'' of the winner, indicated by
its ABC state, or cut its link with the winner and rewired that link by
connecting it to another, randomly selected, node.
All nodes therefore maintained
their initial degree.
Depending on the density of links and the propensity of losing nodes to rewire 
verses changing their opinion, a variety of behavior can occur, 
including both reaching consensus and network fragmentation.     

In our model the system evolves according to the following rules:
From any configuration $\mathcal{C}$,
in a Monte Carlo sweep (MCS), we execute $N$ updates, each of which consists
of choosing a random pair of nodes and, with probability $r$ and $\left(
1-r\right)$, attempt to change the states of, respectively, nodes and links.
Suppose the pair $\left(  i,j\right)$ is chosen. For a node-update, if
$a_{ij}=0$ or if $\alpha_{i}=\alpha_{j}$, then $\mathcal{C}$ remains
unchanged, but a pair with a link connecting it ($a_{ij}=1$) and $\alpha_{i}\neq\alpha_{j}$
(regardless of $\tau_{i,j}$) will be changed according to the rules
(\ref{ABC-rules}), while all links are \textit{unaffected}. Additionally, the
temperament of the winning node in the $ABC$ cyclic competition is inherited
by the `offspring'. For example, when an $AC$ pair turns into $CC$, both of
the nodes at the end have the temperament that the original $C$ node had. On
the other hand, for a link-update, the nodes remain \textit{unchanged}, while
the link value $a_{ij}$ is \textit{assigned}, regardless of its previous state, to
be $1$ or $0$ if the temperaments ($\tau_{i,j}$) are both $E$'s or both $I$'s,
respectively. If $\tau_{i}\neq\tau_{j}$, $a_{ij}$ is set to $1$ or $0$ with
equal probability. This link update rule is not the one introduced in the
original $XIE$ model \cite{LSZepl2012}. Instead, it is in the spirit of
`heat-bath dynamics' that often are used in simulations of the Ising
model~\cite{Derrida1987,Barber1988,Coniglio1989}. Though such a rule does not
lead to the remarkable behavior reported in \cite{LSZepl2012,BLSZpre15}, it
does allow tractable differential equations for the link dynamics to be
written. Such a simplification is justifiable, as this study is an exploration
of systems with node-link co-evolution.

The full stochastic dynamics can -- given the above rules -- be written in 
terms of a master equation for the probability to observe a system configuration $ \mathcal{C}  $ at time $t$, $P\left(  \mathcal{C},t\right)  $, and solved exactly employing a stochastic Gillespie simulation~\cite{Gillespie1976}; see next section. Here, we briefly
discuss some general features of the stochastic dynamics. First, one should note that if the population of one of the node types ever vanishes, then it will be remain zero: there is no mechanism to create a species anew. Let's consider some limiting cases first:

\begin{enumerate}
\item $r=0\,$: There is no node dynamics. With frozen nodes states, the system
reverts simply to the $XIE$-model with fixed number of $I$'s and $E$'s. Only
the $I$-$E$ links fluctuate between being present and absent. The steady state
probability distribution $P^{\ast}\left(  \mathbb{A}\right)  $ is a simple
product of appropriate $p\left(  a_{ij}\right)  $'s.

\item $r=1$: There is no link dynamics. The network is frozen at the initial
topology, on which the final state crucially depends, as the system evolves
with only the $ABC$-dynamics. Depending on the topology of the network interesting phenomena emerge \cite{Szolnoki2014,Masuda2006,Szabo2004}.

\item If the $I$'s go extinct and $r\neq0,1$, then the network will quickly
evolve to a complete graph and the system reverts to the well-mixed $ABC$-model.

\item If instead, the $E$'s go extinct, then all links will eventually
disappear and any $\left\{  \alpha_{i}\right\}  $ is an absorbing state.

\item If any $\alpha \in \{A,B,C\}$ goes extinct, the competition among the remaining two is
trivial, though the final state may again consist of arbitrary ratios provided
they are all introverts (where all links are absent).
\end{enumerate}

From these limits, one aspect is clear: there are many absorbing states for
$r>0$. Here we are not interested in calculating extinction times but in identifying and characterizing non-trivial, long-living,
`quai-stationary' states (QSS), similar as the `active states' in many epidemic
models.

There are only two control parameters, $N$ and $r$, in our model. By contrast, there are many `order parameters,' characterizing the collective behavior of this system. The most natural set consists of $27=6+21$ quantities: $N_{\alpha
,\tau}$, the total number of each node type, as well as $L_{\alpha,\tau
;\alpha^{\prime},\tau^{\prime}}$, the total number of existing links between
individuals of each pair of types. Of course, the set of $N$'s is constrained
by $N=\Sigma_{\alpha,\tau}N_{\alpha,\tau}$. At any given time, not every
individual of type $\left(  \alpha,\tau\right)  $ is connected to every one of
type $\left(  \alpha^{\prime},\tau^{\prime}\right)  $. Therefore
$L_{\alpha,\tau;\alpha^{\prime},\tau^{\prime}}$ is a fluctuating quantity,
even for configurations with fixed $N_{\alpha,\tau}$ and $N_{\alpha^{\prime
},\tau^{\prime}}$. Also, unlike the absence of self-loops ($a_{ii}\equiv0$),
the links between individuals of a single type ($L_{\alpha,\tau;\alpha,\tau}$)
is a non-trivial, dynamic variable (lying in the range $\left[  0,N_{\alpha
,\tau}\left(  N_{\alpha,\tau}-1\right)  /2\right]  $).

\section{Stochastic simulations and mean-field theory}

Using Monte Carlo simulations, our initial explorations began with a random
collection of all six nodes types, with $N$ up to 1000, connected by a random,
half filled network. Not surprisingly, the early stages of the dynamics appear
`chaotic' with 27 intertwined variables. Often, extinction of species sets in quite quickly. However, in many of the simulation runs,
the system settles into relatively regular and long-living QSS. We find that
these QSS contain the `most diverse' set of three node types, namely, all
three species and both temperaments are present. Of course, we may expect this
behavior, given our discussion above on the limiting cases (3-5). Since they
appear to be more tractable, we will devote our attention only to such QSS in
the rest of this paper.'

There are two distinct classes of `diverse' QSS. Exploiting cyclic symmetry,
they can be labeled as $A_{E}B_{E}C_{I}$ or $A_{E}B_{I}C_{I}$. For simplicity,
we will focus here mainly on the former and only comment briefly on the
latter. In either cases, $\alpha$ and $\tau$ are uniquely related and, from
now on, we will drop the $\tau$ label and write, e.g., $N_{A}$ and $L_{BC}$
instead of $N_{A_{E}}$ and $L_{B_{E}C_{I}}$, respectively. To re-emphasize,
our $A$'s and $B$'s are extroverts, while $C$'s are introverts.

\begin{figure}[tbp]
\centering
\includegraphics[width=0.5\textwidth]{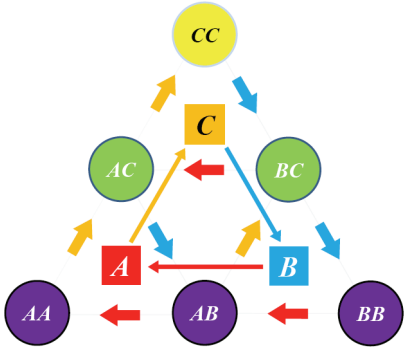}
\caption{ Schematic diagram showing the results of a node update in a
$A_{E}B_{E}C_{I}$ QSS, showing the co-evolutionary nature of the dynamics.
Particles in the square boxes represent individuals
of each node type, while those in circular urns
represent links of the appropriate category. The dark urns (purple-online)
represent E-E links, While the medium (green-online) and light colored
(yellow-online) urns represent I-E and I-I links, respectively. If a node
changes in, say, an $A+B \rightarrow2A$ update, then a particle moves from box
$B$ to box $A$, represented by the long thin dark gray (red-online) arrow. All
the links that the changed node has also change, thus particles moves from
urns $BA$, $BB$, and $BC$ to urns $AA$, $AB$, and $AC$, respectively,
represented by the short thick dark gray (red-online) arrows. The medium gray
(blue-online) and light gray (yellow-online) arrows indicate the particle
moves resulting from $B+C \rightarrow2B$ and $C+A \rightarrow2C$ updates,
respectively. By contrast, during a link update, at most, a single particle
will be added to or removed from the appropriate urn. }
\label{fig1}%
\end{figure}

Of course, the links in our system are non-directional, so that $L_{\alpha
\beta}=L_{\beta\alpha}$. Thus, there are only 3 $N$'s and 6 $L$'s to monitor,
numbers that can be regarded as particles in 3 `square boxes' and 6 `round
urns'; see Fig.~\ref{fig1}. Furthermore, the 3 $N$'s are constrained:
$N=\Sigma_{\alpha}N_{\alpha}$, leaving 8 independent numbers. When a node
changes its state, say through an $A+B\rightarrow2A$ update, a particle moves
from one box to another (denoted by the thin arrows in Fig.~\ref{fig1}). All
the links to this node will also change character, and the corresponding
particles in the urns must move also (denoted by the thick arrows in
Fig.~\ref{fig1}). These considerations will play a key role when we derive 
approximate mean-field equations for $\partial_{t}L_{\alpha\beta}$. By contrast, the
moves associated with a link update are simple: adding or removing a
`particle' in an urn. For example, if an unconnected $AB$ pair is chosen, then
we would add a `particle' to the $L_{AB}$ urn (always, in this case). Note
that the urns on each row are updated with the same rule: Always remove one at
the top ($CC$), always add for urns at the bottom row, and add/remove with
equal probability for those in the middle.

\begin{figure}[ptb]
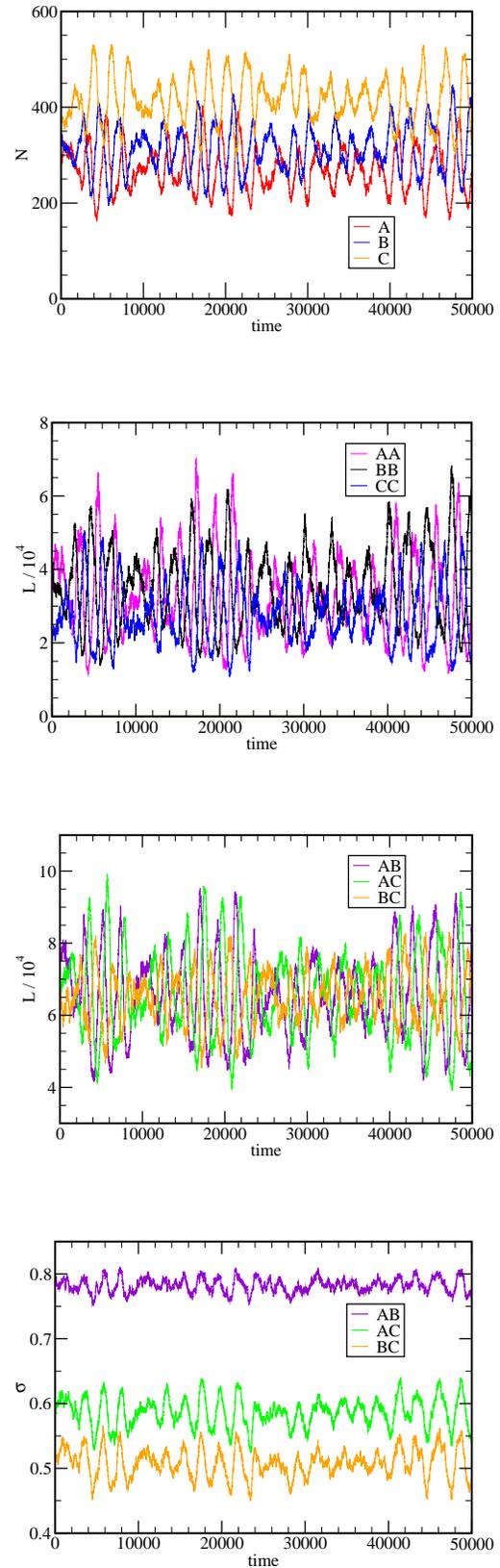

\centering
\includegraphics[width=0.37\textwidth]{fig2a.eps} \newline
\\ ~
\\ ~
\newline
\includegraphics[width=0.37\textwidth]{fig2b.eps} \newline%
\\ ~
\\ ~
\newline
\includegraphics[width=0.37\textwidth]{fig2c.eps} \newline%
\\ ~
\\ ~
\newline
\includegraphics[width=0.37\textwidth]{fig2d.eps} \newline
\caption{ Results from a very
long ($10^{9}$ MCS) run with $N=1000$ and $r = 4/N$. Figures show a typical
$5\times10^{4}$ section of, from top to bottom, a) population numbers of $A$,
$B$, and $C$. b) total self-links $AA$, $BB$, and $CC$. c) total cross-links
$AB$, $AC$, and $BC$. d) fraction of cross-links, which are remarkably in
phase.}%
\label{fig2}%
\end{figure}

To focus on the QSS in simulation studies, various values of ($N,r$) were
chosen. The system was typically initialized near the symmetry point
($N_{\alpha}\cong N/3$) and randomly half-filled with links. It was then
evolved for up to $\thicksim10^{9}$ MCS, during which time the 9 variables
were measured. We first discovered that, for a narrow range of $r$, the life
time of a QSS can be orders of magnitude longer than the typical extinction
time of the $ABC$ model, as much as $O\left(  10^{5}\right)  $ MCS. Encouraged
by the presence of such unexpected longevity, and seeking insight into the
QSS's, we postulated a set of equations for the evolution of the macroscopic
quantities, based on a mean-field approach and discussed below. The fixed
points and their associated stability properties led us to perform further
simulations at $r\cong4/N$. At this value of $r$, runs with $N=1000$ remain
active after $10^{9}$ MCS! Figure~\ref{fig2} illustrates the typical behavior in
such a QSS, more details of which will be discussed below.

Let us first consider the mean-field equations, however, as they provide some
insight into the nature of the co-evolving system. The full equations are
given in the Appendix. Here we only outline their derivation. Formulating how
the node variables change is straightforward. They only change through node
updates, which occur with rate $r$, and during which a chosen link, if
present, will lead to one of its nodes being altered. For example, there are
$L_{AC}$ links connecting an $A$ and a $C$. If one of these links is chosen
and a node update is attempted, then $N_{A}$ will decrease by one and $N_{C}$
will increase by one. Thus, instead of Eq.~\eqref{Neqn}, we may write
\begin{equation}
\partial_{t}N_{A}=r\left(  L_{AB}-L_{AC}\right)  \text{ \ and cyclic.}
\label{NLeqn}%
\end{equation}
Unlike the standard $ABC$-model, the action of the $I$'s means that
$L_{\alpha\beta}<N_{\alpha}N_{\beta}$ typically. Thus, we define a useful
variable,
\begin{equation}
\sigma_{\alpha\beta}\equiv L_{\alpha\beta}/N_{\alpha}N_{\beta} \, ,
\label{sigma}%
\end{equation}
which plays the role of an effective interaction rate in Eq.~\eqref{ABC-rules}.
For example, we can cast the $\partial_{t}N_{A}$ equation above as
$\partial_{t}N_{A}=r\sigma_{AB}N_{A}N_{B}-r\sigma_{AC}N_{A}N_{C}$. Though this
form is neater, the presence of the variables $\sigma$ signals complications,
as the derivation of the link equations below will show. Note that, since the
nodes do not change during link-updates, there are no terms proportional to
$\left(  1-r\right)  $. Of course, the equations for the different node
variables sum to $\partial_{t}N=0$ as $N$ is a constant of motion.

Describing the evolution of the link variables $L_{\alpha\beta}$ is more
involved, as it is affected by both kinds of updates. There are terms due to
link updates, which occur at a rate of $\left(  1-r\right)  $: a chosen link
will be cut or added according to the temperament of its two nodes, and the
number of $I$-$I$ ($E$-$E$) links can only decrease to zero (increase towards
the maximal value), while the $I$-$E$ links are driven towards half of the
maximum value. Thus, we have $\left(  r-1\right)  L_{CC}$ in the $\partial
_{t}L_{CC}$ equation, and e.g., $\left(  r-1\right)  \left\{  L_{AC}%
-N_{A}N_{C}/2\right\}  $ in the $\partial_{t}L_{AC}$ equation, and $\left(
r-1\right)  \left\{  L_{AB}-N_{A}N_{B}\right\}  $ in the $\partial_{t}L_{AB}$
equation. In addition, terms proportional to $r$ are needed to account for
changes induced by a node-update, as shown by the thick arrows in
Fig.~\ref{fig1}. At the microscopic level these terms involve the probability
of 3-node (say, $A$, $B$, and $\gamma$) clusters. Consider choosing a
\textit{particular} connected $AB$ pair, which will change to an $AA$ pair,
under our rules. But then, all the \textit{other} links this $B$ had will also
change, as a $B\gamma$ link is now an $A\gamma$ link, and the 3-node cluster
changes: $A$-$B$-$\gamma\rightarrow A$-$A$-$\gamma$. Denoting the number of
$\gamma$'s linked to $B$ by $k_{B\gamma}$, we see that $L_{A\gamma}%
$/$L_{B\gamma}$ gains/loses by this amount. Unfortunately, a complication is
that $k_{B\gamma}$ differs from one $B$ node to another. In the spirit of mean-field theory, we replace it by the average, namely, $L_{B\gamma}/N_{B}$. Such
standard approximation schemes for moment closure~\cite{Demirel2014} can be
exploited to arrive at a set of terms for the link equations. For example, in
the $\partial_{t}L_{AB}$ equation, we would include gain terms like
$L_{BC}k_{CA}\simeq L_{BC}L_{CA}/N_{C}$ (from choosing the $BC$ pair in a
cluster linked as $B$-$C$-$A$) and loss terms like $-L_{CA}k_{AB}\sim
-L_{AC}L_{AB}/N_{A}$ (from choosing the $CA$ pair in a cluster linked as
$C$-$A$-$B$). The result of such straightforward but tedious considerations
are given as the first set of equations in the Appendix. Together with the set of Eq.~\eqref{NLeqn}, these will be the basis of our theory.

To continue, it is convenient to define the fractions
\begin{subequations}
\begin{align}
\eta_{\alpha}  &  \equiv N_{\alpha}/N\label{eta} \, , \\
\lambda_{\alpha\beta}  &  \equiv2L_{\alpha\beta}/N\left(  N-1\right) \, ,
\label{lamb}%
\end{align}
\end{subequations}
and consider the `thermodynamic limit': $N\rightarrow\infty$ with
\begin{equation}
\rho\equiv r\left(  N-1\right)  /2 \label{rho}%
\end{equation}
fixed. Then rescaling $t$ by $N$, only one control parameter, $\rho$,
remains~\footnote{Note that $\rho\in\left[  0,\left(  N-1\right)  /2\right]  $
and approaches $\left[  0,\infty\right]  $ in the thermodynamic limit. For
simplicity, we keep $\rho$ finite in our numerical analysis and so, we have
$1-r=1-2\rho/N\rightarrow1$ below.}. The full set of mean-field equations for
these variables simplify slightly, as shown in the Appendix. We provide two
examples here:%
\begin{equation}
\partial_{t}\eta_{A}=\rho\left(  \lambda_{AB}-\lambda_{CA}\right) \, , 
\label{etaA}%
\end{equation}%
\begin{align}
\partial_{t}\lambda_{AB}  
&  =\rho\left\{  
\frac{\lambda_{BC}\lambda_{CA}}{\eta_{C}}
{-}\frac{\lambda_{AB}\lambda_{AC}}{\eta_{A}}
{+}2\frac{\lambda_{BB}\lambda_{BA}}{\eta_{B}}
{-}\frac{\lambda_{AB}^{2}}{\eta_{B}}\right\}
\nonumber\\
&  \quad -\left\{  \lambda_{AB}-2\eta_{A}\eta_{B}\right\}  \, .
\label{lamAB}%
\end{align}
Despite these simplifications, it is not feasible to solve these equations
analytically. Even the fixed point equations involve solving non-linear
(algebraic) equations for 5 variables. Before discussing the numerical
results, let us offer some insight into their behavior.

First, note that the fraction of the total number of links, 
$\lambda \equiv \sum_{\alpha}\lambda_{\alpha \alpha}+\sum_{\alpha \neq \beta}\lambda_{\alpha \beta}$,
satisfies
a very simple equation: 
\begin{equation}
\partial_{t}\lambda
=(1-r)\left\{  \eta_{A}+\eta_{B}-\lambda\right\}  
\label{eq:lambda}
\end{equation}
Here we keep $\left(  1-r\right)  $, instead of
writing $\left(  1-2\rho/N\right)  \rightarrow1$, to highlight the absence of
$r$ terms in the evolution of the total link-number, which is conserved during
node-updates. Equation \eqref{eq:lambda} is intuitively reasonable, as
it simply forces $\lambda$ to follow the fraction of extroverts.

\begin{figure}[ptb]
\centering
\includegraphics[width=0.42\textwidth]{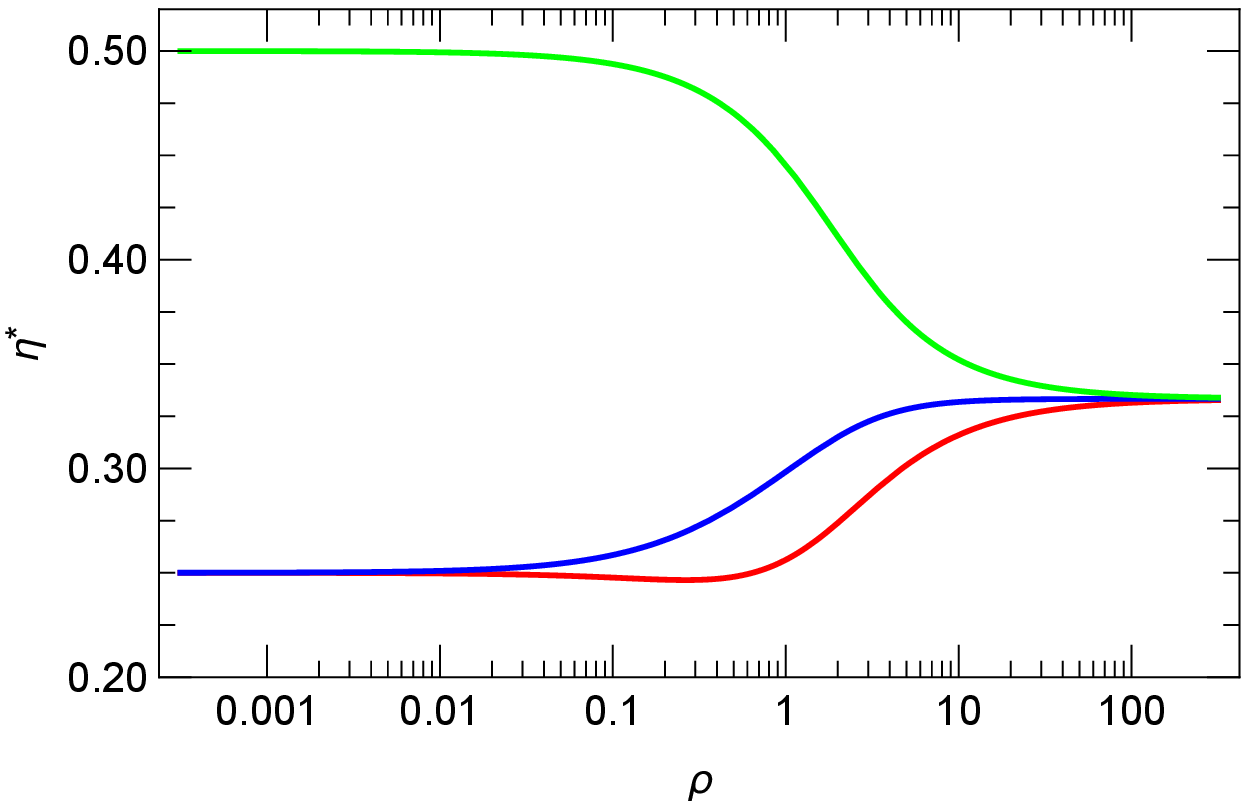} \newline
\newline
\includegraphics[width=0.42\textwidth]{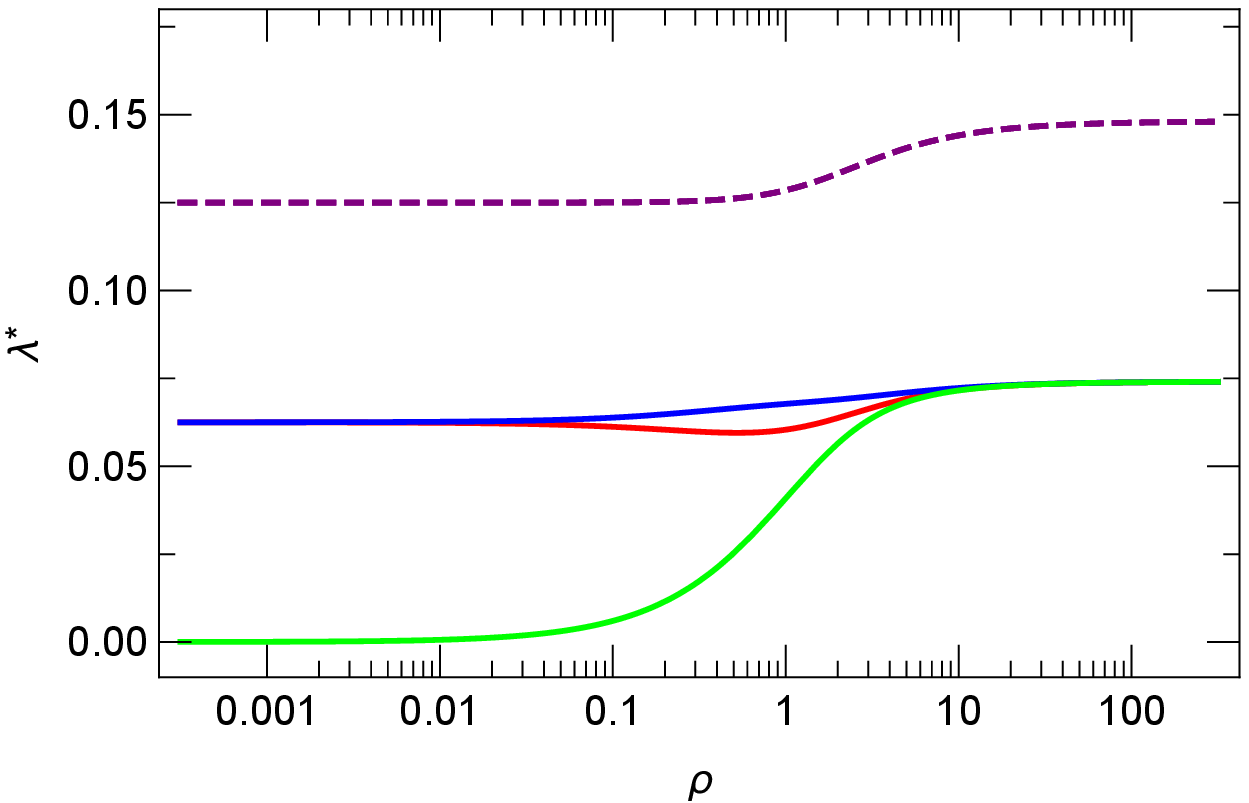} \newline
\newline
\includegraphics[width=0.42\textwidth]{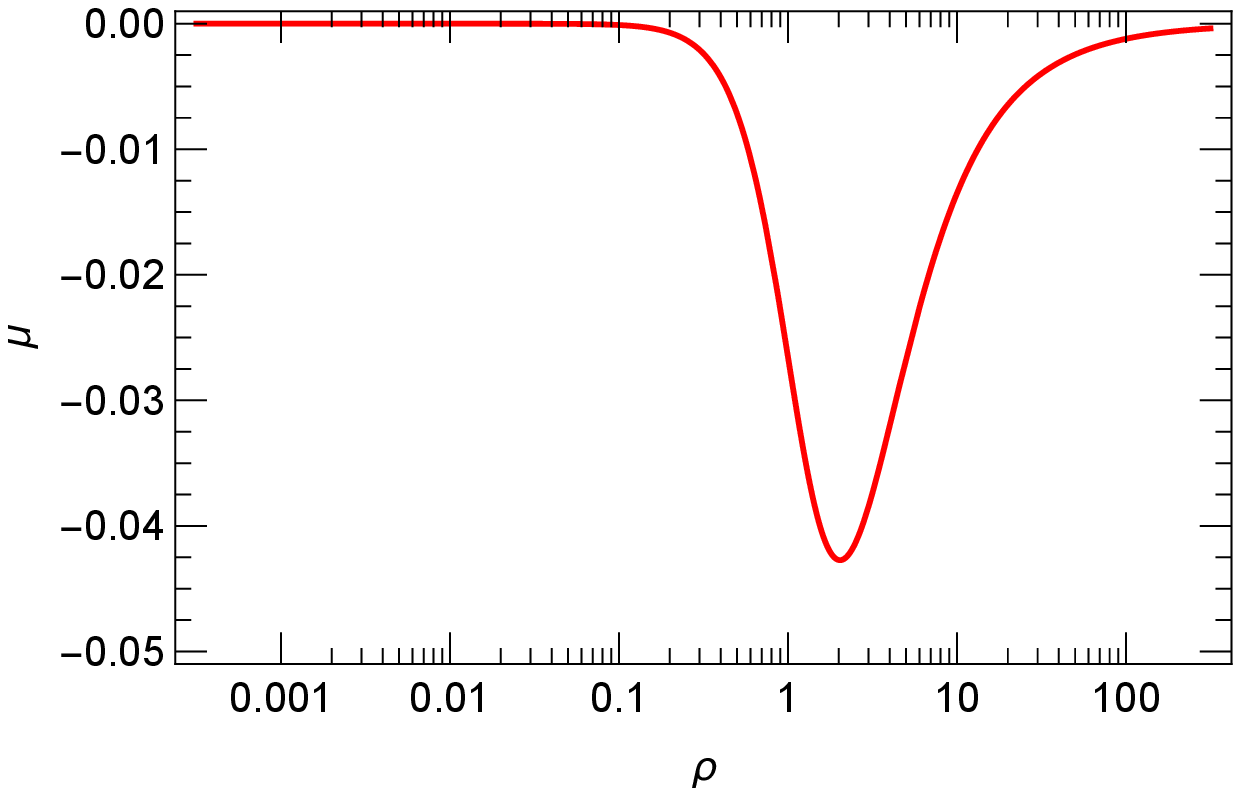} \newline
\caption{ Solution of the mean-field equation and fixed-point stability analysis, as a function of $\rho$.
Top figure:
fixed point values of $\eta$, from top to bottom: C (green online), B
(blue online), A (red online). 
Middle figure: fixed point values of
$\lambda$. Dashed line is for all three types of cross-links. For the
self-links, from top to bottom : BB (blue online), AA (red online), CC (green
online). 
Bottom figure: real
part of the slowest eigenvalue, $\mu$. }%
\label{fig3}%
\end{figure}

As a next step we study the fixed points of the mean-field equations (denoted by a superscript: $^{\ast}$) and their neighborhoods. Despite some simplifications
($\lambda_{AB}^{\ast}\,{=}\,\lambda_{BC}^{\ast}\,{=}\,\lambda_{CA}^{\ast}\,{\equiv}\, x^{\ast}$, $\lambda^{\ast}\,{=}\,\eta_{A}^{\ast}{+}\eta_{B}^{\ast}$), the analytic form of
$(  \eta_{\alpha}^{\ast},\lambda_{\alpha\beta}^{\ast})  $ is quite
complex (as they are solutions of non-linear equations for 5 variables). Let
us discuss briefly the simpler $r\rightarrow0,1$ limits. For the former
($r\rightarrow0$), the rapid link dynamics leads to effective rates of $1/2$
for all $I$-$E$ pairs, so that we have $\eta_{A,B}^{\ast}\,{=}\,\eta_{C}^{\ast
}/2=1/4$, $x^{\ast}\,{=}\,1/8$, $\lambda_{AA,BB}^{\ast}\,{=}\,1/16$, and $\lambda_{CC}^{\ast}\,{=}\,0$. For $r\rightarrow1$, the fast $ABC$ dynamics is expected to symmetrize all variables, so that we have $\eta_{A,B,C}^{\ast}\,{=}\,1/3$. Meanwhile, $\lambda_{\alpha\alpha}^{\ast} \,{=}\, x^{\ast}/2$. To obtain a definite result, we appeal to $3\left( \lambda_{\alpha\alpha}^{\ast}+x^{\ast}\right) \,{=}\, \lambda^{\ast} \,{=}\, \eta_{A}^{\ast}+\eta_{B}^{\ast} \,{=}\, 2/3$ and arrive
at$\ x^{\ast} \,{=}\, 4/27$ and $\lambda_{\alpha\alpha}^{\ast} \,{=}\, 2/27$. Apart from these
limiting cases, we can find $(\eta_{\alpha}^{\ast},\lambda_{\alpha
\beta}^{\ast})  $ numerically. They are shown, as a function of $\rho$,
in Figs.~\ref{fig3}A and B. For finite $N$, these values differ slightly,
for example, by $\thicksim 1\%$ for $N = 1000$.

Linearizing the evolution equations around the fixed point, we find the
following remarkable properties in the spectrum of the stability matrix. There
are typically multiple complex conjugate pairs, the real parts of all
eigenvalues are negative, so that the fixed point is always \emph{stable}!
Except for one complex conjugate pair, the magnitudes of these real parts are
$O\left(  1\right)  $, leading to rapid decay of six modes. Meanwhile, the
real part of the last complex conjugate pair, $\mu$, is $O\left(
10^{-2}\right)  $ or less, so they are associated with the dominant modes at
late times. Plotting $\mu$ in Fig.~\ref{fig3}C, we find $\mu\rightarrow0$ at
the two limits, indicating that the fixed point becomes marginally stable, as
expected. There is also a surprisingly significant dip around $\rho \,{=}\, 2$. At this $\rho$, the fixed point values are%
\begin{subequations}
\begin{align}
\eta_{A,B,C}^{\ast}  &  \cong~0.274,~0.315,~0.411 \, \\
\lambda_{AA,BB,CC}^{\ast}  &  \cong~0.0637,~0.0690,~0.0565 \, \\
x^{\ast}  &  \equiv\lambda_{\alpha\neq\beta}^{\ast}\cong0.133 \, .
\end{align}
\end{subequations}
The eigenvalues associated with the two slowest modes are $\mu \,{\cong}\, {-}0.0427 {\pm} 1.43i$. 
Meanwhile, the next eigenvalue is $-0.853$, so that the next mode
decays about $20$ times faster. Thus, we can expect the dominant stochastic
evolution to take place in a `slow plane' spanned by the eigenvectors of the
two slowest modes. The direction of these eigenvectors in phase space encode valuable
information. We find the magnitudes of \emph{all} the components are comparable.
Thus, all quantities ($\eta,\lambda$) vary with similar amplitudes. The
relative phases of the three $\eta$ components are quite close to
$\pm120^{\circ}$, reflecting the underlying three-fold cyclic behavior.
Meanwhile, the intra-species links ($\lambda_{\alpha\alpha}$) and the
cross-links ($\lambda_{\alpha\beta}$) are essentially in phase with
$\eta_{\alpha}$ and $\eta_{\alpha}\eta_{\beta}$, respectively. A significant
consequence of the latter is that, even if the amplitudes in $\eta$ and
$\lambda$ can be large, the ratios $\sigma_{\alpha\beta}$ oscillate with much
smaller amplitudes.

The main conclusion of this linear stability analysis is as follows. The
trajectory of our system in 8-dimensional space relaxes quickly onto a `slow
plane.' In the absence of noise, the phase portrait within this plane is that of a stable spiral such that the dynamics converges to the fixed point $(  \eta_{\alpha}^{\ast},\lambda_{\alpha\beta}^{\ast})  $, similar as an underdamped, simple oscillator. Thus, we observe the \textit{stochastic} evolution to be well approximated by noisy oscillations (see Fig.~\ref{fig2}). Since fluctuations move the system away
from the attractive fixed point, the long time behavior consists of
oscillations with random amplitudes (within a finite range, of course). The dominant
frequency is given by the imaginary part of the eigenvalue ($1.43$ here) and
comparable to $\rho$ ($2$ here).

All these properties are consistent with the data from simulations (see
Fig.~\ref{fig2}), confirming that these mean-field equations capture the
typical behavior of the stochastic system. The cross-links, $L_{\alpha\beta}$,
oscillate considerably, around the same value with phase lags of
$\thicksim120^{\circ}$ (Fig.~\ref{fig2}c). The main contributions to these
variations are, however, due to the time-dependent, total number of
available links: $N_{\alpha}N_{\beta}$. When these are factored out, by
considering $\sigma_{\alpha\beta}$ instead, we find oscillations which are not
only much smaller in amplitudes but also approximately in phase
(Fig.~\ref{fig2}d).
Furthermore, these effective interaction rates, when normalized by
$\hat{\sigma}\equiv\sigma/\Sigma_{\alpha\beta}\sigma_{\alpha\beta}$, are
almost constant, varying by at most $5\%$. These constant values
\begin{equation}
\hat{\sigma}_{AB,AC,BC}\cong~0.418,~0.312,~0.270
\end{equation}
are entirely consistent with the $\eta_{\alpha}^{\ast}$'s~\cite{RMF06,BRSFprl09}.

\begin{figure}[ptb]
\centering
\includegraphics[width=0.45\textwidth]{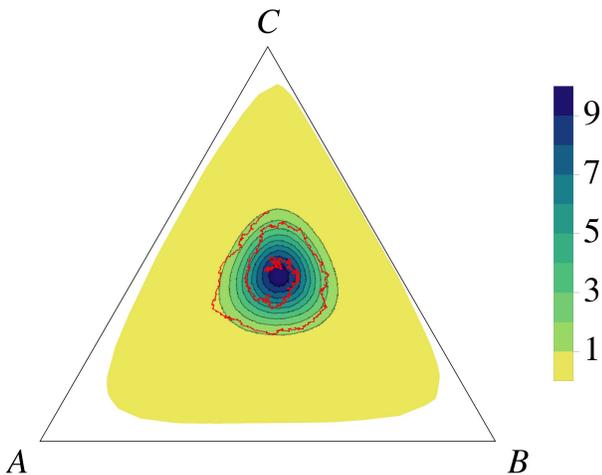}\caption{ Contour
plot of the frequencies of the occurrence of $(A,B,C)$ during a very long
($10^{9}$ MCS) run, representing (a section of) the quasi-stationary
distribution, PQSS. The units in the legend are arbitrary; total number of
occurrences being the length of the run: $10^{9}$. Also shown is a typical
trajectory (red online) of length 5000 MCS starting near the center at
t=90,022,001. }%
\label{fig4}%
\end{figure}

Another perspective of the QSS is the distribution function, $P^{QSS}\left(
\eta_{\alpha},\lambda_{\alpha\beta}\right)  $, which can be obtained by
compiling a histogram with the long time series. Figure~\ref{fig4} shows only
the section of $P^{QSS}$ associated with the node variables. Not surprisingly,
the central contours are essentially circular, where $P^{QSS}$ is well
approximated by a symmetric Gaussian. To illustrate the stochastic nature of
the evolution, we also included a short section of the trajectory (red
online). This short trajectory begins near the center and spirals away. Later,
not shown, it spirals back toward the fixed point.

Unfortunately, while the mathematical analysis shows how the addition of link
dynamics and social temperament diversity promotes the coexistence of the
species, an intuitively clear picture remains elusive. Here, let us offer a
few observations. First, the slight asymmetry in the fixed point values of
$\eta_{\alpha}^{\ast}$ can be argued as follows. Once an $A$ is converted to a
$C$, it starts cutting its links, an in particular those to $B$. This action
implies that it can survive longer than otherwise. Another perspective is that
a smaller fraction of $C$'s participate in the game, leading to a larger total
population. Meanwhile, since they `consume' the $A$'s and `feed' the $B$'s,
their abundance can account for the slight difference between $A$ and $B$.
Next, let us focus on the $A$'s, with the smallest average population. As they
near extinction, the are more likely to recover than in the ordinary $ABC$
model, since their predators tend to cut links, thereby reducing those
effective interactions. Indeed, in simulations a strong correlation between
$\eta_{A}\ll1$ and rapid decrease in the effective rate $\sigma_{CA}$ is
observed, without significant variations in $\sigma_{AB}$, the rate they prey
on the $B$'s. A similar, but `opposite,' scenario appears to be the case for
the $B$'s. As they near extinction, their effective predation rate,
$\sigma_{BC}$, increases, while their susceptibility to the $A$'s,
$\sigma_{AB}$, remain roughly constant. In this sense, the drift towards
extinction in the standard $ABC$ model is averted by the action of the introverts.

Finally, we report very similar findings for the $A_{E}B_{I}C_{I}$ case.
Changes to equations such as (\ref{etaA}) and (\ref{lambAB}) are minor, only
in the $\left(  1-r\right)  $ terms in the $\partial_{t}\lambda_{B\alpha}$
equations. The most stable system occurs at a slightly different $\rho$, but
is much more stable ($\mu\thicksim-0.3$). Simulation results are also similar,
except that the system are extremely stable, showing few sustained
oscillations as fluctuations decay rapidly. Of course, the scenarios painted
above must be modified. 

\section{Summary and outlook}

To summarize, in an attempt to merge two standard paradigms in statistical
physics, we introduced a simple model to explore co-evolution of node and link
degrees of freedom. Specifically, we combined the cyclically competing game of
three species with the link dynamics in a population of extreme introverts and
extroverts. Whereas the former always ends in extinction within a short time,
we find that the addition of link dynamics leads to a surprising long lived
state of coexistence when both introverts and extroverts are present. Though a
mean field approach, analyzed numerically, provides much insight into this new
state, an \textit{intuitive} understanding of how co-evolution and diversity
prevent extinction remains elusive.

There are many avenues for further study of this particular model, in addition
to the QSS in the complementary $A_{E}B_{I}C_{I}$ case. Interesting issues
include incorporating noise into the mean-field equations, analytic
perturbative treatments near the two extremes of $r$, paths of partial
population collapse from 6 to 3 node types, possible existence of QSS's
involving $4$ or $5$ types, intrinsically asymmetric interspecies interaction
rates, potential emergence of phase transitions between extinction and
coexistence, and symmetry in co-evolutionary dynamics near such
transitions~\cite{HRB2014}. Beyond our simple model, many generalizations come
to mind readily. One is to impose a structure in the interaction network, such
as making it a spatial network~\cite{Penrose2003,NGB2015}. Also, the link
dynamics of the original $XIE$ model leads to an extreme Thouless
effect~\cite{BLSZpre15}. If such a dynamics were implemented, instead of the heatbath dynamics used here, can we expect similar novel phenomena?

Of course, our $ABC$-$XIE$ model is just a prototype, designed only for a
theoretical exploration of the non-equilibrium statistical physics of
co-evolution. More realistic complex systems in nature should be considered,
an example being how susceptible individuals would naturally reduce contacts
in the presence of an epidemic. In this sense, an SIS model with adaptive
networks~\cite{GDB2006,SS2008,JLSZ2012} mark
the inception of a serious pursuit. Recent work studying competing bacteria
colonies that produce and degrade antibiotics are another
example~\cite{KishonyNature2015}. 

\bigskip

\begin{acknowledgments}
We thank C.I. del Genio, J.J. Dong, T. Gross, J. Knebel, L.B. Shaw, and Z.
Toroczkai for illuminating discussions, and especially G\"{u}ven Demirel
during the initial phases of this project. This research is supported in part
by the US National Science Foundation, through grant DMR-1507371 (KEB and RKPZ). E.F. acknowledges support by the German Excellence Initiative via the program `NanoSystems Initiative Munich' (NIM). 
\end{acknowledgments}

\cleardoublepage

\normalem
\bibliographystyle{apsrev4-1}
\bibliography{ABC_XIE}

\cleardoublepage

\section{Appendix}

Here, we provide some details of the equations for the evolution of
$L_{\alpha\beta}$ . The terms stemming from link updates, being proportional
to $\left(  1-r\right)  $, are easily understandable. Those needed to account
for a node update are more involved; see thick arrows in Fig.~\ref{fig1}. In
the main text, we discussed how the leading terms associated with the
probability of 3-node clusters arise. However, there is no unique scheme for
the next leading terms. Here, we choose a set which reduces properly to the original
$ABC$ model (a fully connected population) in the limit of $r=1$. The
result is \begin{widetext}
\begin{equation}
\partial _{t}L_{AA}=r\left\{ \frac{L_{AB}L_{AB}}{N_{B}}-2\frac{L_{AA}L_{AC}}{%
N_{A}}-\frac{L_{AC}+L_{AB}}{2}\right\} -(1-r)\left\{ L_{AA}-\frac{%
N_{A}\left( N_{A}-1\right) }{2}\right\}  \label{AA}
\end{equation}%
\begin{equation}
\partial _{t}L_{BB}=r\left\{ \frac{L_{BC}L_{BC}}{N_{C}}-2\frac{L_{BB}L_{BA}}{%
N_{B}}-\frac{L_{BC}+L_{BA}}{2}\right\} -(1-r)\left\{ L_{BB}-\frac{%
N_{B}\left( N_{B}-1\right) }{2}\right\}  \label{BB}
\end{equation}%
\begin{equation}
\partial _{t}L_{CC}=r\left\{ \frac{L_{CA}L_{CA}}{N_{A}}-2\frac{L_{CC}L_{CB}}{%
N_{C}}-\frac{L_{CB}+L_{CA}}{2}\right\} -(1-r)\left\{ L_{CC}\right\}
\label{CC}
\end{equation}%
\begin{equation}
\partial _{t}L_{AB}=r\left\{ \frac{L_{BC}L_{CA}}{N_{C}}-\frac{L_{AB}L_{AC}}{%
N_{A}}+2\frac{L_{BB}L_{BA}}{N_{B}}-\frac{L_{AB}L_{AB}}{N_{B}}+L_{AB}\right\}
-(1-r)\left\{ L_{AB}-N_{A}N_{B}\right\}  \label{AB}
\end{equation}%
\begin{equation}
\partial _{t}L_{CA}=r\left\{ \frac{L_{CB}L_{BA}}{N_{B}}-\frac{L_{CA}L_{CB}}{%
N_{C}}+2\frac{L_{AA}L_{AC}}{N_{A}}-\frac{L_{CA}L_{AC}}{N_{A}}+L_{CA}\right\}
-(1-r)\left\{ L_{CA}-\frac{N_{C}N_{A}}{2}\right\}  \label{AC}
\end{equation}%
\begin{equation}
\partial _{t}L_{BC}=r\left\{ \frac{L_{CA}L_{AB}}{N_{A}}-\frac{L_{CB}L_{BA}}{%
N_{B}}+2\frac{L_{CC}L_{CB}}{N_{C}}-\frac{L_{CB}L_{CB}}{N_{C}}+L_{CB}\right\}
-(1-r)\left\{ L_{CB}-\frac{N_{C}N_{B}}{2}\right\}  \label{BC}
\end{equation}%
\end{widetext}It is straightforward to verify that setting $r$ to unity and
substituting $L_{\alpha\alpha}=N_{\alpha}\left(  N_{\alpha}-1\right)  /2$ and
$L_{\alpha\beta}=N_{\alpha}N_{\beta\neq\alpha}$ into Eqns. (\ref{AA}-\ref{BC})
gives us the same set as Eqs.~(\ref{Neqn}).

Keeping these next-to-leading order terms, the total number of links,
$\mathcal{L\equiv\Sigma}_{\mathcal{\alpha}}L_{\alpha\alpha}+\mathcal{\Sigma
}_{\mathcal{\alpha\neq\beta}}L_{\alpha\beta}$, satisfies the simple equation%
\begin{equation}
\partial_{t}\mathcal{L}=(1-r)\left\{  \left(  N-1\right)  \left(
N-N_{C}\right)  /2-\mathcal{L}\right\}
\end{equation}
As for fixed points, the following results are trivial: $L_{AB}^{\ast}%
=L_{BC}^{\ast}=L_{CA}^{\ast}\equiv X^{\ast}$, $\mathcal{L}^{\ast}=$ $\left(
N-1\right)  \left(  N-N_{C}^{\ast}\right)  /2$. Thus, the number of
independent variables reduces to five. Nevertheless, the analytic forms of
$\left(  N_{\alpha}^{\ast},L_{\alpha\beta}^{\ast}\right)  $ are, in general,
quite complex. The exception are, as in the main text, the $r\rightarrow0,1$
limits. Here, we provide the next-to-leading terms as well. For the former
($r\rightarrow0$), we have $N_{A,B}^{\ast}=N_{C}^{\ast}/2=N/4$, $X^{\ast
}=N^{2}/16$, $L_{AA,BB}^{\ast}=N\left(  N-4\right)  /32$, and $L_{CC}^{\ast
}=0$. For $r\rightarrow1$, the fast $ABC$ dynamics is expected to symmetrize
all variables, so that we have $N_{A,B,C}^{\ast}=N/3$. Exploiting
$\mathcal{L}^{\ast}=$ $\left(  N-1\right)  \left(  N-N_{C}^{\ast}\right)  /2$,
the final results are $X^{\ast}=2N\left(  N+1\right)  /27$ and $L_{\alpha
\alpha}^{\ast}=N\left(  N-7/2\right)  /27$.

Finally, let us present the full set of equations for the fractions, in the
thermodynamic limit: \begin{widetext}
\begin{equation}
\partial _{t}\eta _{A}=\rho \left( \lambda _{AB}-\lambda _{CA}\right)
;~~\partial _{t}\eta _{B}=\rho \left( \lambda _{BC}-\lambda _{AB}\right)
;~~\partial _{t}\eta _{C}=\rho \left( \lambda _{AC}-\lambda _{BC}\right)
\label{etaAB}
\end{equation}%
\begin{equation}
\partial _{t}\lambda _{AA}=\rho \left\{ \frac{\lambda _{AB}^{2}}{\eta _{B}}-2%
\frac{\lambda _{AA}\lambda _{AC}}{\eta _{A}}\right\} -\left\{ \lambda
_{AA}-\eta _{A}^{2}\right\}  \label{lambAA}
\end{equation}%
\begin{equation}
\partial _{t}\lambda _{BB}=\rho \left\{ \frac{\lambda _{BC}^{2}}{\eta _{C}}-2%
\frac{\lambda _{BB}\lambda _{BA}}{\eta _{B}}\right\} -\left\{ \lambda
_{BB}-\eta _{B}^{2}\right\}  \label{lambBB}
\end{equation}%
\begin{equation}
\partial _{t}\lambda _{CC}=\rho \left\{ \frac{\lambda _{CA}^{2}}{\eta _{A}}-2%
\frac{\lambda _{CC}\lambda _{CB}}{\eta _{C}}\right\} -\lambda _{CC}
\label{lambCC}
\end{equation}%
\begin{equation}
\partial _{t}\lambda _{AB}=\rho \left\{ \frac{\lambda _{BC}\lambda _{CA}}{%
\eta _{C}}-\frac{\lambda _{AB}\lambda _{AC}}{\eta _{A}}+2\frac{\lambda
_{BB}\lambda _{BA}}{\eta _{B}}-\frac{\lambda _{AB}^{2}}{\eta _{B}}\right\}
-\left\{ \lambda _{AB}-2\eta _{A}\eta _{B}\right\}  \label{lambAB}
\end{equation}%
\begin{equation}
\partial _{t}\lambda _{CA}=\rho \left\{ \frac{\lambda _{CB}\lambda _{BA}}{%
\eta _{B}}-\frac{\lambda _{CA}\lambda _{CB}}{\eta _{C}}+2\frac{\lambda
_{AA}\lambda _{AC}}{\eta _{A}}-\frac{\lambda _{CA}^{2}}{\eta _{A}}\right\}
-\left\{ \lambda _{CA}-\eta _{C}\eta _{A}\right\}  \label{lambAC}
\end{equation}%
\begin{equation}
\partial _{t}\lambda _{CB}=\rho \left\{ \frac{\lambda _{CA}\lambda _{AB}}{%
\eta _{A}}-\frac{\lambda _{CB}\lambda _{BA}}{\eta _{B}}+2\frac{\lambda
_{CC}\lambda _{CB}}{\eta _{C}}-\frac{\lambda _{CB}^{2}}{\eta _{C}}\right\}
-\left\{ \lambda _{CB}-\eta _{C}\eta _{B}\right\}  \label{lambBC}
\end{equation}%
\end{widetext}
where, of course, $\eta_{C}$ stands for $1-\eta_{A}-\eta_{B}$. We should
emphasize that, in this limit, the period of typical oscillations is of the
order of $1/\rho$, since they are controlled by the $ABC$ dynamics.

\end{document}